\documentclass[runningheads]{llncs}
\usepackage[T1]{fontenc}
\usepackage{graphicx}
\newcommand{\eat}[1]{}

\newcommand{\pset}{\mathscr{P}}

\newcommand{\alphabet}{\Sigma}

\newcommand{\loc}{\mathrm{loc}}

\newcommand{\dcset}[1]{{\downarrow}#1}

\newcommand{\Da}{{\Downarrow}}

\newcommand{\gtrn}{\Rightarrow}

\newcommand{\traces}[1][\Sigma]{%
    \mathrm{Tr}^*(#1)%
}
\newcommand{\inftraces}[1][]{%
         \ifthenelse{\isempty{#1}}{\mathrm{Tr}^\omega(\alphabet)}{\mathrm{Tr}^\omega(#1)}%
}
\newcommand{\maxtraces}[1][]{%
	\ifthenelse{\isempty{#1}}{\mathrm{Tr}^\mathrm{max}(\alphabet)}{\mathrm{Tr}^\mathrm{max}(#1)}%
}
\newcommand{\alltraces}[1][]{%
         \ifthenelse{\isempty{#1}}{\mathrm{Tr}(\alphabet)}{\mathrm{Tr}(#1)}%
}

\let\phi\varphi

\newcommand{\configs}[1]{C_{#1}}

\newcommand{\view}[2]{{\partial}_{#1}(#2)}

\usepackage{tikz}
\usetikzlibrary{arrows,automata,positioning}

\tikzset{
  state/.style = {draw, circle
  },
  event/.style = {draw, rectangle  },
}

\usetikzlibrary{decorations.pathmorphing} 
\usetikzlibrary{calc}

\definecolor{turquoise}{RGB}{64,224,208}
\usepackage{ amssymb }
\usepackage{mathrsfs} 
\usepackage{mathtools,amscd}
\usepackage{textcomp}
\usepackage{mathrsfs}
\usepackage{rotating}
\usepackage{enumitem}
\usetikzlibrary{positioning}

\begin{document}
\title{Non-deterministic asynchronous automata games and their undecidability
}
%
%
\author{Bharat Adsul\inst{1}\and
Nehul Jain\inst{1}\orcidID{0000-0003-1531-6780} }
\authorrunning{F. Author et al.}
%
\institute{Indian Institute of Technology Bombay, India }
\maketitle              
\begin{abstract}
We propose a new model of a distributed game, called an ATS game,
which is played on a non-deterministic asynchronous transition system –
a natural distributed finite-state device working on Mazurkiewicz traces.
This new partial-information game is played between an environment and
a distributed system comprising of multiple processes.

A distributed strategy 
uses causal past to make the next move. The key algorithmic question is to solve the game, that
is, to decide the existence of a distributed winning strategy. 
 It
turns out ATS games are equivalent to asynchronous games which are known to
be undecidable. We prove that ATS games are undecidable.


\keywords{Mazurkiewicz Traces \and Model of Concurrency \and  Distributed
Synthesis \and Game-theoretic models \and Asynchronous Automata  }
\end{abstract}
%
%
%

The design and analysis of concurrent programs/protocols have always been a challenging task.
Given the importance of concurrent/distributed applications and their pervasiveness, it is 
only natural to ask if a class of such protocols can be synthesized automatically starting
from its specification.

The initial work \cite{PR90}
proposed a setting for distributed synthesis and showed
that it is in general undecidable. In this setting, the processes have access to only 
{\em local} information and are also allowed to communicate only a limited amount of information.
In contrast, recent works \cite{GastinLZ04PS,MadhusudanTY05,Muscholl09,Muscholl15,GenestGMW12,GenestGMW13,Fin14,Gimbert16,Gimbert17}
have allowed 
the processes to access the entire {\em causal past}. The work \cite{GastinLZ04PS} introduced distributed
controller synthesis problem for series-parallel systems and showed that controlled reachability is 
decidable. In \cite{MadhusudanTY05}, the authors consider the case of connectedly communicating processes, that is,
a setting where if two processes do not communicate for a long enough time, they never communicate. It is
shown that the MSO theory in this setting is decidable.

Now we briefly talk about the two lines of work, namely, Petri games and asynchronous games.
Petri games \cite{Fin14} are distributed games played on Petri nets where tokens are considered as players.
Places of the underlying Petri net are divided into system places and environment places. In 
a distributed strategy, a token residing in a system place decides based on its causal history
which of the transitions it will allow next. Several interesting classes of Petri games \cite{Fin14}
have been shown to be decidable. However, it has been recently shown \cite{Fin21} that Petri games
with global winning conditions are undecidable even with two system places and one environment place.

Asynchronous/control games from \cite{GenestGMW12,GenestGMW13} are played on deterministic 
asynchronous automata. This setting involves a fixed number of processes. The set of actions in
which these processes participate is partitioned into controllable and uncontrollable actions.
In a distributed/control strategy, each process decides based on its casual past, the next set
of control actions that it wants to allow. In \cite{GenestGMW12,GenestGMW13}, it has been
shown asynchronous games over acyclic architectures are decidable. The work \cite{Gimbert16,Gimbert17} introduces
a general class of asynchronous games, called decomposable games which include games from
\cite{GastinLZ04PS,MadhusudanTY05,GenestGMW12,GenestGMW13}, and proves their decidability. 
Another recent exciting result from \cite{Gimbert21} shows that asynchronous games with
local reachability objectives is undecidable even with six processes.

In this work, we propose a new model of distributed games, called ATS games, which are played on 
{\em non-deterministic} {\bf a}synchronous {\bf t}ransition {\bf s}ystems. Such a game is played
between an environment and a distributed system comprising of a fixed number of processes. 
An environment is best seen as a {\em scheduler} of actions: it can schedule purely local actions --
modelling, for instance, user-requests at a particular process, or it can schedule synchronizing
actions -- modelling communication between different processes. 
Once an action is scheduled, the processes participating in that action are free to share
their individual causal past with each other and based on their collective causal past, they decide
how to advance their joint-state by selecting an appropriate transition of the underlying non-deterministic
asynchronous transition system. These games are equipped with a regular trace-based winning condition. The primary question addressed in this work is-`Does there exist a winning strategy in the game?'

In Section~\ref{sec:game} we introduce the model as required and introduce winning conditions.

In  Section~\ref{sec:eqvi} 
we see that it turns out these games are harder than Asynchronous games. This makes these games in general undecidable.

\section{ATS Game}\label{sec:game}




In general a game consists of a \emph{game arena}, a \emph{start position} in the arena and a \emph{winning condition}. \emph{Two players} play the game. 
The moves possible for each player is given by the game arena. The winning condition decides which player wins 
a particular play.
A play is a sequence of consecutive moves by the players that is allowed in the arena.

The environment is a singular entity and has complete 
information of an ongoing play. On the other hand the system consists of a set of processes. Each process 
has only its causal past in its knowledge in an ongoing play or interaction. 
 When two 
processes interact they share their causal past which is the maximum amount of 
information they  can gather restricted by the structure of the game.

Let $\mathcal{P}=\{1,2,...,K\}$ be a set of processes. A distributed alphabet over $\mathcal{P}$ is a family 
$\widetilde{\Sigma} = (\Sigma_i)_{i \in \mathcal{P}}$. Let $\Sigma = \bigcup_{i \in \mathcal{P}} \Sigma_i$. For $a \in \Sigma$,
we set $loc(a) = \{i \in \mathcal{P}~| a \in \Sigma_i\}$. By $(\Sigma, I)$ we denote the corresponding trace alphabet, i.e., $I$
is the independence relation $\{(a,b) \in \Sigma^2~| loc(A) \cap loc(b) = \emptyset \}$. The corresponding dependence relation $(\Sigma \times \Sigma) \backslash I$ is denoted by D.


\subsection{Trace}

\begin{definition}
A trace over $\widetilde{\Sigma} = (\Sigma_i)_{i \in \mathcal{P}}$ is a labeled partial order $F = (E,\leq, \lambda)$
where,
  \begin{itemize}
  \item 
  $E$ is a (possibly infinite)set of events
  \item
  $\lambda: E \mapsto \Sigma$ is a labeling function 
  \item $\leq$ is a partial order on $E$ satisfying

    \begin{itemize}
    \item $(\lambda(e),\lambda(e')) \in D$ implies $e \leq e'$ or $e' \leq e$
    
    
    \item $e\lessdot f$ implies $(\lambda(e),\lambda(f))\in D$, where

    by definition $e \lessdot e'$ iff $e \leq e'$ and for each $e''$ with $e \leq e'' \leq e'$,
    either $e'' = e$ or $e'' = e'$. 
    
    \item $\forall e \in E, \downarrow\! \!e$ is finite, where
    $\downarrow\! \!e = \downarrow\! \!\{e\}$. 
    \end{itemize}
  \end{itemize}
  
\end{definition}

For $X \subseteq E$, we let 
$\downarrow\! \!X   = \{ y \in E ~|~ y \leq x \text{ for some } x \in X\}$.
For $e \in E$ we set $\downarrow\! \!e = \downarrow\! \!\{e\}$. Going forward $loc(e)$ abbreviates $loc(\lambda(e))$.

Let $TR(\widetilde{\Sigma})$ denote the set of all traces over $\widetilde{\Sigma}$.
Henceforth a trace means a trace over $\widetilde{\Sigma}$ unless otherwise mentioned.
A trace language is simply a subset of $TR(\widetilde{\Sigma})$.

\subsection {Configuration}

A subset $c \subseteq E$ is a configuration iff $\downarrow\! \!c = c$.We let $\mathcal{C}_F$ denote the set of configurations
of trace $F$. Notice that $\emptyset$ and $E$ are configurations.
More importantly $\downarrow\! \!e$ is a configuration for every $e \in E$. Also, $\downarrow\! \!e - e$ is
a configuration for every $e \in E$.

Let $c \in \mathcal{C}_F$  and $i \in \mathcal{P}$. 
The  i-view of $c$  is the best configuration that the process $i$ is aware
of when $c$ has occurred. 
We write ${}^i(c)$ to denote $c \cap E_i$ where $E_i =\{ e \in E~|~ i \in loc(e)\}$.
Then $\downarrow\!\!{}^i(c)$ is the i-view of $c$ and is defined as  $\downarrow\!\!{}^i(c) = \downarrow\!\!(c \cap E_i)$.
We note that $\downarrow\!\!{}^i (c)$ is also a configuration.
It is easy to see that if $\downarrow\!\!{}^i (c) \neq \emptyset$ then there exists $e \in E_i$ such that
$\downarrow\!\!{}^i (c) = \downarrow\!\! e$. 
All $i$ events in a configuration $c$ are strictly ordered, as  these events, all depend on each other.
Therefore, the last event of $c$ in which process $i$ has participated is uniquely defined. We denote this as $max_i(c)$.
$max_i(c)=\{e\in c ~|~ i\in loc(e), \forall e'\in c i\in loc(e') \text{ implies } e'\leq e\}$
So, $\downarrow\!\!{}^i (c) = \downarrow\!\!max_i(c)$. 
Also $maxset(c) = \bigcup_{p \in \mathcal{P}} max_p(c)$. 
For  $F = (E,\leq, \lambda)$ we write $F$ as a set of events instead of $E$ when context is clear.

For $P \subseteq \mathcal{P}$ and $c \in \mathcal{C}_F$, we let $\downarrow\!\!{}^P (c)$ denote the set 
$\bigcup \{\downarrow\!\!{}^i (c) ~| i \in P \}$. Once again, $\downarrow\!\!{}^P (c)$ is a configuration. 
It represents the collective knowledge of the processes in $P$ about $c$.

Let $F= (E,\leq,\lambda) \in TR(\widetilde{\Sigma})$. There are two natural transition relations that one may associate with F.
The event based transition relation $\Rightarrow_F \subseteq \mathcal{C}_F \times E \times \mathcal{C}_F$ is defined as 
$c \stackrel{e}{\Longrightarrow_F} c'$
iff $e \notin c$ and $c \cup \{e\} = c'$. The action based transition relation 
$\rightarrow_F \subseteq \mathcal{C}_F \times \Sigma \times \mathcal{C}_F$ is defined as 
$c \xrightarrow{a}_F c'$
iff there exists $e \in E$ such that $\lambda(e) = a$ and $c \stackrel{e}{\Longrightarrow_F} c'$.

To talk about a game, we define automata over a set of processes.
\subsection{Asynchronous transition system}
\label{ASS}

We equip each process $i \in \mathcal{P}$ with a finite non empty set  of local $i$-states denoted
$S_i$. We set $S= \bigcup_{i \in \mathcal{P}}S_i$ and call $S$ the set of local states.
We let $P$ range over non empty subsets of $\mathcal{P}$ and let $i,j$ range over $\mathcal{P}$. 
Each $P$-state, $s$ is a map $s: P\rightarrow S$ such that $s(j) \in S_j$ for every $j \in P$.
We let $S_P$ denote the set of P states. We 
call $S_\mathcal{P}$ as the set of global states.

If $P' \subset P$ and $s \in S_P$ then $s(P')$ is the restriction to $P'$. We use $a$ to abbreviate $
loc(a)$ when talking about states. Then $a$-state is just a $loc(a)$ state and $S_a$ denotes the set of all 
$loc(a)$ states. If $loc(a) \subseteq P$ and $s$ is a $P$-state we shall write $s_a$ to mean $s_{loc(a)}$.

\begin{definition}
An asynchronous transition system over distributed alphabet $\widetilde{\Sigma}$
is a tuple
$\mathcal{A} = \langle \{S_p\}_{p \in \mathcal{P}},  \{\delta_{a}\}_{a \in \Sigma} \rangle$
where, 
  \begin{itemize}
  \item $\mathcal{P} = \{1,2,...,K \}$ is a set  of $K$ processes.
  Each process $p$ has local states given by set $S_p$.
   
  \item 
  For each action there is a transition relation 
  $\delta_a \subseteq S_a \times S_a$

 \end{itemize}
  
\end{definition}

An $a$-move  involves only those process that are local to action $a$ i.e., $loc(a)$. 
We construct  global transition relation
$\widehat{\delta_a}\subseteq S_{\mathcal{P}}\times S_{\mathcal{P}}$ for each action $a$ from 
$\delta_a \subseteq S_a\times S_a$ as follows.
$\widehat{\delta_a}(s) \ni s'$ iff $ \delta_a (s_a) \ni s'_a$ and $s_{\mathcal{P}-loc(a)} = s'_{\mathcal{P}-loc(a)} $. An 
action a is enabled in $s$ $a \in enabled(s)$ if $\widehat{\delta_a}(s)$ is not empty $\widehat{\delta_a}(s)\neq \emptyset$.

A trace run of $\mathcal{A}$ over $F \in TR(\widetilde{\Sigma})$ is $(F,\rho)$ where,
$\rho$ is a map $\rho:\mathcal{C}_F \rightarrow S_{\mathcal{P}}$
such that $\rho(\emptyset) = s_{in}$ and for every 
$c \xrightarrow{a}_F c'$, 
$(\rho(c),\rho(c'))\in \widehat{\delta}_a $. 
There can be multiple runs for a single trace.

\subsection{Strategy}
Now we define the crucial notion of a distributed strategy for $G$. For the sake of brevity, we assume
that every action is enabled at every global state of the underlying transition system $A$. We abbreviate
this assumption by saying that $A$ (or $G$) is complete.

\begin{definition}A distributed strategy for $G$ is a function $ \sigma: \traces \to S$
where
\begin{itemize}
	
	\item $\sigma: \traces \to S$ maps {\em finite} traces
 	to global states of $A$ so that $\sigma(\emptyset) = s_0$ and, for all
	$t \in \traces, a \in \alphabet: (\sigma(t), \sigma(ta)) \in\; \gtrn_a$.
\end{itemize}
\end{definition}
Due to the completeness assumption on $A$, the environment
can `play/schedule' every action at every trace.
A distributed strategy records through the function $\sigma$ a (deterministic) response to these
actions; when the environment's choice corresponds to event $e$
(at global state $\sigma(\Da e)$), the participating processes have access to complete causal
past $\dcset e$ in order to advance the global state from $\sigma(\Da e)$ to $\sigma(\dcset e)$.

A play $(t, \rho)$ of $G$ is said to conform to $\sigma$ if for all $c \in \configs{t}$,
$\rho(c) = \sigma(c)$. A distributed strategy $\sigma$ is said to be winning if all maximal plays
conforming it are won by the distributed system. 
In other words, a winning distributed strategy ensures that no matter
what scheduling choices are made by the environment, the resulting plays played according
to it are always won by the distributed system.

\subsection{Game Definition}

\begin{definition}

A game  is of the form $\mathcal{G} = (\mathcal{A}, s_0, \text{Win})$
where 
\begin{itemize}
 \item $\mathcal{A} = \langle \{S_p\}_{p \in \mathcal{P}},  \{\delta_{a}\}_{a \in \Sigma} \rangle$ is 
 an asynchronous transition system.
 \item $s_0 \in \mathbb{S} $ is  an initial global state, where a play of the game begins.
 \item $\text{Win}$  is a set of plays in which the system is winning.
\end{itemize}

\end{definition}

For a $\mathcal{P}$-state $s $, $enabled(s) = \{ a \in \Sigma ~|~ \exists s' \in S ~(s,s') \in \delta_a \}$

We define a  game using a transition system  $\mathcal{A}$ , as $\mathcal{G} = (\mathcal{A}, s_{0} ,\text{Win})$, where  
$\text{Win} \subseteq plays$ is the  set of plays where system wins.
State $s_{0} \in S$ is an initial state, where a play of the game begins.

A game begins at the initial state $s_0$. Environment plays an enabled action at the current state. System chooses the next state consistent with the transition system. The two player play in an alternating fashion.

This sequence of moves gives rise to a play where each move is consistent with the asynchronous automaton $\mathcal{A}$.
Each play can be represented as a trace run of this automaton.
A play $(t,\rho:\mathcal{C}_t \rightarrow S_{\mathcal{P}})$ is called winning(for the system) when 
$(t,\rho) \in \text{Win}$.

A play $(t,\rho:\mathcal{C}_t \rightarrow S_{\mathcal{P}})$ is called losing (winning for the environment) when 
$(t,\rho) \not\in \text{Win}$.

A maximal play is a play which cannot be extended by either player. 
A finite play is said to be partial if one of the players has an enabled move.

Intuitively a player's {\bf strategy} for a player is something that tells a player
where to move next. In general a player's strategy may decide the next move based on the entire history which is the partial 
play upto the last move made by either player. Therefore a natural way of representing a strategy is using a function that maps 
each possible history to an enabled next move.
A play conforms to a strategy for the system 
if each move is made according to the strategy. Similar statement holds for the environment.
A strategy for a player is winning if all maximal plays conforming to the strategy are winning.

{ \em
The control problem for  $\mathcal{G} = (\mathcal{A},\text{Win})$, 
is to determine from which states  there exists a winning strategy for the system. This set of states is called the winning region 
denoted $W_{sys}$.
}


{ \em
The decision problem for  
$\mathcal{G} = (\mathcal{A}, s_0, \mathbb{S})$
is to determine if there exists a winning strategy for the 
system from $s_0$.

}

\begin{example}

We consider a game  illustrated in  Figure \ref{fig:det}.
Consider a set of processes $\{q,r\}$.$a$ and $b$ are local actions for process $q$ and $r$ respectively.i.e.$\Sigma_q = \{a\}$ and $\Sigma_r = \{b\}$. The transitions on $a$ are $q_0 \xrightarrow{a} q_1$ and $q_0 \xrightarrow{a} q_2$. The transitions on $b$ are $r_0 \xrightarrow{b} r_1$ and $r_0 \xrightarrow{b} r_2$. The initial state is $(q_0,r_0)$, and red states are the only  unsafe states.

\label{ex:rollback}

A sequential strategy exists for System starting at $(q_0,r_0)$ to avoid unsafe states given by deterministic transitions,
$(q_0,r_0)\xrightarrow{b}(q_0,r_1)$,
$(q_0,r_0)\xrightarrow{a}(q_2,r_0)$,
$(q_0,r_1)\xrightarrow{a}(q_1,r_1)$ and
$(q_2,r_0)\xrightarrow{b}(q_2,r_2)$.
By determinacy of sequential games, environment cannot have a winning strategy
from this state.
Environment cannot have a winning strategy
even in this distributed game.

In this game however on action $a$ let $q$ choose $q_0\rightarrow_a q_1$ move, and on $b$  $r$ choose $r_0\rightarrow_b r_1$ move.
As actions $a$ and $b$  occur on disjoint sets of processes a transition on action $a$ by $q$ 
does not enable or disable an action $b$ at $r_0$ in $r$, neither does a transition on $b$ 
changes whether action $a$ is enabled or not. In fact $q$ would not know whether an action $b$ 
has been chosen by environment and $r$ made its move. So if environment chooses actions $b$ 
followed by $a$, the choices of system can not vary.

Therefore, there are four possible distributed strategies for the system. For each we can 
give a trace for which system will lose.

\begin{tabular}{cc}

  System strategy    & environment move\\
  top-right    & ac\\
  top-left      &ac\\
 bottom-right  &abc\\
  bottom-left  & bc
\end{tabular}

We see that although there is a sequential strategy there is no distributed strategy. In fact, the environment wins in all maximal plays.

    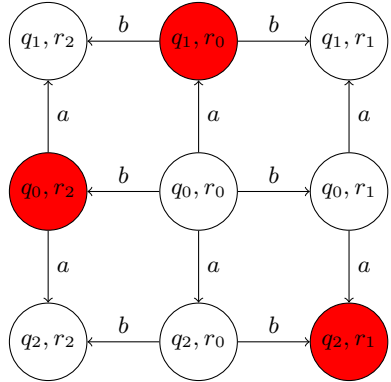
\begin{figure}
    \begin{center}

        \begin{tikzpicture}[scale=1,     
]

\node[state](12) at (0,4) {$q_1,r_2$} ;
\node[state,fill=red](10) at (2,4) {$q_1,r_0$} ;
\node[state](11) at (4,4) {$q_1,r_1$} ;
\node[state,fill=red](02) at (0,2) {$q_0,r_2$} ;
\node[state](00) at (2,2) {$q_0,r_0$} ;
\node[state](01) at (4,2) {$q_0,r_1$} ;
\node[state](22) at (0,0) {$q_2,r_2$} ;
\node[state](20) at (2,0) {$q_2,r_0$} ;
\node[state,fill=red](21) at (4,0) {$q_2,r_1$} ;

\path[->]
(10) edge node[above]{$b$} (12)
(10) edge node[above]{$b$} (11)

(00) edge node[above]{$b$} (02)
(00) edge node[above]{$b$} (01)

(20) edge node[above]{$b$} (22)
(20) edge node[above]{$b$} (21)

(02) edge node[right]{$a$} (12)
(02) edge node[right]{$a$} (22)

(00) edge node[right]{$a$} (10)
(00) edge node[right]{$a$} (20)

(01) edge node[right]{$a$} (11)
(01) edge node[right]{$a$} (21)

;
    \end{tikzpicture}

\caption{\emph{ Example:} 
 $\mathcal{P}=\{q,r\}$, $\Sigma_q=\{a\}$, $\Sigma_r=\{b\}$, $s^{in}=(0,0)$, $\mathbb{S} = \{(q_1,r_2),(q_1,r_1),(q_0,r_0),(q_0,r_1),(q_2,r_2),(q_2,r_0)\}$
} \label{fig:det}
  \end{center}
   \end{figure}

\end{example}

\section{Comparision with Asynchronous games}\label{sec:eqvi}
Asynchronous games have been extensively studied
 \cite{Gimbert21}  and are recognized as undecidable. Moreover, ATS games exhibit equivalence to Asynchronous games, meaning that for every ATS game, an equivalent asynchronous game can be constructed. This equivalence ensures that if a system possesses a winning strategy in one type of game, it will possess it in the other as well. Conversely, an ATS game can be constructed for every asynchronous game, maintaining this symmetry.

This parity underscores the undecidability of ATS games. Furthermore, the undecidability extends to 6-process Asynchronous games, implying the same for 6-process ATS games. Further examination reveals that games featuring 3 decision makers alongside 3 additional deterministic processes suffice to solve the Post's Correspondence Problem (PCP), a problem known to be undecidable. However, the decidability of 2CDM (Two Central Decision Makers) games remains unknown.

In this article, we will delve into the undecidability of asynchronous transition systems (ATS) games by reducing any asynchronous game to an ATS game.

In an asynchronous game \cite{GenestGMW12}, a distributed alphabet, denoted as $\Pi$, is partitioned into two sets of actions: controllable actions, denoted as $\Sigma^{sys}$, and uncontrollable actions, denoted as $\Sigma^{env}$. This partitioning reflects the distinct roles played by the system and the environment in influencing the game dynamics.

Central to the study of asynchronous games is the concept of a deterministic automaton, represented as $B = (Q, \xRightarrow{a})$, where $Q$ is the set of states and $\xRightarrow{a}$ denotes the transition relation for each action $a$. On a play (i.e., a sequence of actions taken during the game), the state of the automaton is uniquely determined by the corresponding trace of actions.

\eat{
We write $\Plays_p(\Aa)$ for the set of plays that are
$p$-views:
\[\Plays_p(\Aa)=\set{\view_p(u) \mid u\in\Plays(\Aa)}\,.
\]

A \emph{strategy for a process} $p$ is a function
$\s_p:\Plays_p(\Aa)\to 2^{\Ssys_p}$, where $\Ssys_p=\set{a \in \Ssys
\mid p\in\loc(a)}$.  We require in addition, for every $u \in
\Plays_p(\Aa)$, that $\s_p(u)$ is a subset of the actions that are
possible in the $p$-state reached on $u$. A \emph{strategy} is a
family of strategies $\set{\s_p}_{p\in \PP}$, one for each process.
}

To effectively navigate an asynchronous game, players employ strategies. A strategy , for each process, consists of a mapping that associates each trace to a "set of actions". Specifically, if a trace $u$ is consistent with the strategy and $ua$ results in a valid run in $B$, then the following conditions hold:

    If $a \in \Sigma^{sys}$: For $ua$ to be consistent with the strategy, all processes participating in action $a$ must map it to a "set containing $a$``. This ensures that the system has control over $a$, and its execution aligns with the strategy.

    If $a \in \Sigma^{env}$: $ua$ is consistent with the strategy. This allows the environment to exercise its uncontrollable influence over $a$ while adhering to the defined strategy.

The key insight is that the undecidability of asynchronous games can be extended to ATS games, making them equally complex and challenging to analyze. This implies that reasoning about optimal strategies and outcomes in ATS games is inherently intractable, and there exists no general algorithm to determine the best course of action for the players.

Recall that the decidability question for both games is- "does there exist a winning strategy? ``
In the following lemma, we observe that the undecidability of asynchronous games implies the undecidability of ATS games.

\begin{lemma}
 Asyn $<$ ATS:  For any state based winning condition given an Asynchronous game $G$ we can construct an ATS game $G'$ with the same number of processes and same order of actions.
\end{lemma}

%



\paragraph{Construction:}  
Given an asynchronous game \( G = (S_i, \delta_{a}, \Sigma = \Sigma^{sys} \cup \Sigma^{env}, Win) \), we construct an ATS game \( G' = (Q_i, \xRightarrow{a}, \Pi, Win') \) as follows:

\begin{enumerate}
    \item \textbf{States:} The set of states \( Q_i \) for any process consists of both pure states \( S_i \) and states augmented with a set of actions that the process can choose in that state, i.e., \( S_i \times 2^{\Sigma_i} \). Therefore,  
    \[
    Q_i = S_i \cup (S_i \times 2^{\Sigma_i}).
    \]

    \item \textbf{Actions:} The actions \( \Pi \) consist of both deterministic actions \( \Sigma \) and non-deterministic actions, called choice actions \( c_i \) for each process \( i \). These choice actions are enabled at every state to allow the system to select which actions are permissible:
    \[
    \Pi = \Sigma \cup \{ c_i \mid i \in \pset \}.
    \]
    The location of each choice action \( c_i \) is given by \( \loc(c_i) = \{i\} \).

    \item \textbf{Transitions for Choice Actions:} For each \( c_i \in \Pi \), the transitions are defined as follows:
    \[
    s_i \xrightarrow{c_i} (s_i, X \subseteq \Sigma_i),
    \]
    where \( X \) contains all environment actions enabled in \( s_i \). In a pure state, only this choice action is allowed, enabling the system to select its controllable actions without interfering with uncontrollable ones.

    \item \textbf{Transitions for Regular Actions:} Once the processes have chosen their action sets, the next state is determined by the chosen actions. For \( a \in \Sigma \) where \( \loc(a) = \{i_1, i_2, \dots\} \), we add the transition:
    \[
    [(s_{i_1}, X_{i_1}), (s_{i_2}, X_{i_2}), \dots] \xRightarrow{a} \delta_a(s_a),
    \]
    if \( a \in X_{i_1} \cap X_{i_2} \cap \dots \). This allows the environment to choose the next action among all available options from the system and uncontrollable actions.

    \item \textbf{Play correspondence:} A play \( t \) in \( G \) corresponds to a unique play \( (t', \rho) \) in \( G' \), where \( t' \) is constructed by interleaving choice actions with regular actions. For example, if \( t = abc \), then \( t' = c_a a c_b b c_c c \), where \( c_a = c_{i_1} c_{i_2} \dots \) for \( \loc(a) = \{i_1, i_2, \dots\} \).
    
      \item \textbf{Winning Conditions:} 
      The winning set \( Win' \) in the ATS game is defined as the set of plays corresponding to the winning set \( Win \) in the asynchronous game.
\end{enumerate}

\paragraph{Natural Strategy in \( G' \) Derived from \( G \):}  
Given a strategy \( \rho_i \) in \( G \), we construct a corresponding strategy \( \rho' \) in \( G' \) by selecting the next set of allowed actions according to \( \rho \):
   \[
   \rho'(t'c_i) = \rho_i(t).
   \]

\paragraph{Natural Strategy in \( G \) Derived from \( G' \):}  
Conversely, given a strategy \( \rho' \) in \( G' \), we construct a corresponding strategy \( \rho_i \) in \( G \) by selecting the next set of allowed actions based on \( \rho' \):
   \[
   \rho_i(t) = \rho'(t'c_i)_i.
   \]

\paragraph{Conservation of Winner in strategies:} Due to the one to one correspondence of plays if a strategy is winning in a game the corresponding strategy is winning in the corresponding game.


\paragraph{Undecidability:}  
The special case of a safety winning condition for six processes is known to be undecidable for asynchronous games. By defining the safety winning condition for the ATS game using the same set of unsafe states, we obtain an equivalent undecidable game.

\textbf{Conclusion}

In conclusion, the undecidability of asynchronous games and their equivalence to ATS games highlight the profound complexity of distributed systems and the formidable obstacles faced when attempting to devise optimal strategies in such scenarios. 

It is not the case that the MSO theory of every finite asynchronous transition system is decidable. Hence an interesting question is: what is the precise sub-
class of finite asynchronous transition systems for which the MSO theory is
decidable? Therefore we are keen to reanalyse the undecidability proof in detail with our model to discover further incites into the cause of undecidability.





%
 \bibliographystyle{splncs04}
 \bibliography{main}
%






\end{document}